\documentclass[11pt]{article}

\usepackage{amsmath}
\usepackage{amssymb}
\usepackage{rotating}
\newcommand{\be}{\begin{equation}}
\newcommand{\ee}{\end{equation}}

\begin{document}
\parskip 0pt
\parindent 0pt
\begin{center}
\bf \Large Statistical randomization test for QCD intermittency in
a single-event distribution\footnote{written April 02, 2004; To be
published} \rm \vskip .2in \vskip .5in

 \large \rm  LEIF E.\ PETERSON
\\
\vskip .5in


\textit{Department of Medicine\\ Baylor College of Medicine\\ One
Baylor Plaza, ST-924\\ Houston, Texas, 77030 USA}\\

\vskip .5in

\end{center}

\large
\parindent 0pt

\large
 \textbf{Abstract}
\vskip .1in

A randomization test was developed to determine the statistical
significance of QCD intermittency in single-event distributions.
A total of 96 simulated intermittent distributions based on
standard normal Gaussian distributions of size N=500, 1000, 1500,
2000, 4000, 8000, 16000, and 32000 containing induced holes and
spikes were tested for intermittency.  Non-intermittent null distributions were also
simulated as part of the test.  A log-linear model was
developed to simultaneously test the significance of fit
coefficients for the $y$-intercept and slope contribution to
$\ln(F_2)$ vs.\ $\ln(M)$ from both the intermittent and null
distributions. Statistical power was also assessed for each fit
coefficient to reflect the proportion of times out of 1000 tests
each coefficient was statistically significant, given the induced
effect size and sample size of the Gaussians.  Results indicate
that the slope of $\ln(F_2)$ vs.\ $\ln(M)$ for intermittent
distributions increased with decreasing sample size, due to
artificially-induced holes occurring in sparse histograms.  For intermittent
Gaussians with 4000 variates, there was approximately 70\% power to detect
a slope difference of 0.02 between intermittent and null distributions.
For sample sizes of 8000 and greater, there was more than 70\% power to detect a
slope difference of 0.01.  The randomization test performed
satisfactorily since the power of the test for intermittency
decreased with decreasing sample size.  Power was near-zero when the test
was applied to null distributions.  The randomization test can be used
to establish the statistical significance of intermittency in empirical
single-event Gaussian distributions.

\vskip 0.05in


\parindent 10pt
\parskip 0in
\section{Introduction}

Intermittency has been studied in a variety of forms including
non-gaussian tails of distributions in turbulent fluid and heat
transport [1,2], spikes and holes in QCD rapidity distributions
[3], $1/f$ flicker noise in electrical components [4], period
doubling and tangent bifurcations [5], and fractals and long-range
correlations in DNA sequences [6,7]. The QCD formalism for
intermittency was introduced by Bialas and Peschanski for
understanding spikes and holes in rapidity distributions, which
were unexpected and difficult to explain with conventional models
[3,8,9]. This formalism led to the study of distributions which
are discontinuous in the limit of very high resolution with
spectacular features represented by a genuine physical (dynamical)
effect rather than statistical fluctuation [10-19].

The majority of large QCD experiments performed to date typically
sampled millions of events (distributions) to measure
single-particle kinematic variables (i.e., rapidity, tranverse
momentum, azimuthal angle) or event-shape variables [20-24]. These
studies have employed either the horizontal scaled factorial
moment (SFM)
\be
F_{q} \equiv  \frac{1}{E} \sum_{e=1}^E \frac{1}{M} \sum_{m=1}^M \frac{n_{me}(n_{me}-1)\cdots(n_{me}-q+1)} {\left( \frac{N_e}{M} \right)^{[q]} } , \ee which
normalizes with event-specific bin average over all bins ($N_e/M$),
the vertical SFM
\be
F_{q} \equiv  \frac{1}{M} \sum_{m=1}^M \frac{1}{E} \sum_{e=1}^E \frac{n_{me}(n_{me}-1)\cdots(n_{me}-q+1)} {\left( \frac{N_m}{E} \right)^{[q]} } , \ee
which normalizes with bin-specific average over all events ($N_m/M$),
or the mixed SFM
\be
F_{q} \equiv  \frac{1}{ME} \sum_{m=1}^M \sum_{e=1}^E
\frac{n_{me}(n_{me}-1)\cdots(n_{me}-q+1)} {\left( \frac{N}{ME}
\right)^{[q]}} , \ee which normalizes by the grand mean for all
bin counts from all events ($N/EM$). In the above equations,
$n_{me}$ is the particle multiplicity in bin $m$ for event $e$,
$M$ is the total number of bins, $N_e=\sum_m^M n_{me}$,
$N_m=\sum_e^E n_{me}$, and $N$ is the sum of bin counts for all
bins and events.  A full description of SFMs described above can
be found in [10-12].  There are occassions, however, when only a
single event is available for which the degree of intermittency is
desired.  In such cases, the mean multiplicity in the sample of
events cannot be determined and the normalization must be based on
the single-event SFM defined as
\be
F_{q} \equiv \frac{1}{M} \sum_{m=1}^M
\frac{n_{m}(n_m-1)\cdots(n_m-q+1) }{\left( \frac{N}{M} \right)
^{[q]}} , \ee where $n_m$ is the number of bin counts within bin
$m$, and $N=\sum_{m}n_m$ is the total number of counts for the
single-event.  When intermittency is present in the distribution,
$F_q$ will be proportional to $M$ according to the power-law \be
F_q \propto M^{\nu_q}, \quad \quad M \rightarrow \infty \ee (or
$F_q \propto \delta y ^{-\nu_q}, \delta y \rightarrow 0$) where
$\nu_q$ is the \textit{intermittency exponent}. By introducing a
proportionality constant $A$ into (5) and taking the natural
logarithm we obtain the line-slope formula
 \be \ln(F_q) = \nu_q \ln(M) + \ln(A).
\ee If there is no intermittency in the distribution then
$\ln(F_q)$ will be independent from $\ln(M)$ with slope $\nu_q$
equal to zero and $\ln(F_q)$ equal to the constant term $\ln(A)$.
Another important consideration is that if $\nu_q$ is
non-vanishing in the limit $\delta y \rightarrow 0$ then the
distribution is discontinuous and should reveal an unusually rich
structure of spikes and holes.

The goal of this study was to develop a randomization test to
determine the statistical significance of QCD intermittency in
intermittent single-event distributions.  Application of the randomization
test involved simulation of an
intermittent single-event distribution, multiple simulations of
non-intermittent null distributions, and a permutation-based
log-linear fit method for $\ln(F_2)$ vs.\ $\ln(M)$ for assessing
significance of individual fit coefficients.  The statistical
power, or probability of being statistically significant as a
function of coefficient effect size, Gaussian sample size, and
level of significance, was determined by repeatedly simulating
each intermittent single-event distribution and performing the
randomization test 1000 times. The proportion of randomizations
tests that were significant out of the 1000 tests reflected the
statistical power of each fit coefficient to detect its relevant slope or
y-intercept value given the induced intermittency.

\section{Methods}

\subsection{Sequential steps}
A summary of steps taken for determining statistical power for
each of the log-linear fit coefficients is as follows.

\begin{enumerate}
\item Simulate an intermittent single-event Gaussian distribution of sample size $N$ by
introducing a hole in the interval $(y, y + \Delta_h)_h$ and
creating a spike by adding hole data to existing data in the
interval  $(y, y - \Delta_s)_s$.  (This simulated intermittent
distribution can be replaced with an empirical distribution for
which the presence of intermittency is in question).  Determine
the minimum, $y_{min}$, maximum, $y_{max}$, and range
$\Delta_y=y_{max}-y_{min}$ of the $N$ variates.

\item Assume an experimental measurement error (standard deviation) such as
$\epsilon=0.01$.  Determine the histogram bin counts
$n(m)_{\epsilon}$ for the intermittent distribution using
$M_{max}=\Delta y /\epsilon$ equally-spaced non-overlapping bins
of width $\delta y = \epsilon$.  Apply kernel density estimation
(KDE) to obtain a smooth function (pdf) of the histogram
containing $M_{max}$ bins.  KDE only needs be performed once when
bin width $\delta y = \epsilon$.

\item Use the rejection method based on the pdf obtained from KDE to
simulate a non-intermittent Gaussian.  Determine histogram bin
counts $n(m)_{null,\epsilon}$ over the same range $\Delta y$ of
the intermittent distribution.

\item For each value of $M$, collapse together the histogram bins
determined at the experimental resolution and determine $\ln(F_2)$
and $\ln(M)$ at equal values of $M$ for both the intermittent and
null distribution. Bin collapsing should be started with the first
bin $n(1)$.

\item Perform a log-linear
fit using values of $\ln(F_2)$ vs.\ $\ln(M)$ from both the
intermittent non-intermittent null distributions.  Determine the
significance for each coefficient using the Wald statistic,
$Z_j=\beta_j/s.e.(\beta_j)$ ($j=1,2,...,4$) where $Z_j$ is
standard normal distributed.  For the $bth$ permutation ($b=1,2,...,B)$ where $B=10$,
permute the group labels (intermittent vs.\ non-intermittent) in data records, refit,
and calculate $Z_j^{(b)}$.  After the $B$ permutations, determine
the number of times $|Z_j^{(b)}|$ exceeded $|Z|$.

\item  Repeat steps 4 and 5, only this time start collapsing bins at the
2nd bin $n(2)$.  This essentially keeps $\Delta y$ the same but
shifts the interval over which bin collapsing is performed to
$(y_{min} + \epsilon, y_{max} + \epsilon)$.

\item Repeat steps 3 to 6 a total of 10 times simulating
a non-intermittent null distribution each time, and determine the
significance of each fit coefficient with permutation-based fits.
Thus far, we have used multiple simulations combined with a
permutation-based log-linear fit to determine the statistical
significance of each fit coefficient for one intermittent
distribution (simulated in step 1).  These procedures comprise a
single randomization test to determine significance of fit
parameters for the single-event distribution.

\item To estimate statistical power for each fit coefficient, repeat the randomization test in steps 1-7 1000 times, each time simulating a
new intermittent single-event distribution with the same sample
size and spike and hole intervals.  The statistical power for each
fit coefficient is based on the bookkeeping to track the number of
times $|Z_j^{(b)}|$ exceeds $|Z_j|$ within the permutation-based
log-linear fits.  Report the average values of $\beta_j$ and
$s.e.(\beta_j)$ obtained before permuting group labels to reflect
effect size, and report the statistical power of each fit
coefficient for detecting the induced effect.  This power
calculation step is unique to this study for assessing the
proportion of randomization tests during which each coefficient
was statistically significant.

\end{enumerate}

\subsection{Inducing spikes and holes in Gaussian distributions}
An artifical hole was induced by removing bin counts in the range
$y + \Delta_h$ and placing them into the interval $y - \Delta_s$
as shown in Figure 1. Figures 2 and 3 show the simulated
intermittent Gaussian distributions based on 10,000 variates and
plots of $\ln(F_2)$ vs.\ $\ln(M)$ for $y=2.0$ and $y=0.2$,
respectively. Figure 2a shows the formation of a
hole above and spike below $y=2$ as the hole width $2 + \Delta_h$
increased and spike width $2 - \Delta_s$ decreased. The pattern in
Figure 2b suggests that the level of intermittency based on the
slope of $\ln(F_2)$ vs.\ $\ln(M)$ increases with increasing ratio
$\Delta_h/\Delta_s$. The positive non-zero slopes of $\ln(F_2)$
vs.\ $\ln(M)$ in Figure 2b suggests intermittency at a level
beyond random fluctuation of bin counts.  Figure 3 shows similar
results but for a hole introduced in the range $0.2 + \Delta_h$
and spike in the range $0.2 - \Delta_s$. One can notice in Figure
3a that because of the larger bin counts near $y=0.2$, the
resulting spikes are larger for the same hole sizes presented in
Figure 2a. This supports the rule of thumb that spikes and holes
near the bulk of a Gaussian contribute more to $F_2$.  The overall
point is that the slope of the line $\ln(F_2)$ vs.\ $\ln(M)$
increases with increasing size of the induced hole, decreasing
spike width, and increasing value of the pdf into which hole data
are piled.

\subsection{Histogram generation at the assumed experimental resolution}
It was observed that $F_2$ increased rapidly when the bin size was
smaller than $\delta y =0.01$.  This was most likely due to
round-off error during histogram generation.  Because round-off
error at high resolution can create artificial holes and spikes in
the data, the parameter $\epsilon$ was introduced to represent a
nominal level of imprecision in the intermittent data, which was
assumed to be 0.01.  Thus, the smallest value of $\delta y$ used
was $\epsilon$, at which the greatest number of bins
$M_{max}=\Delta y / \epsilon$ occurred.  In addition, $F_q$ was
only calculated for $q=2$ in order to avoid increased sensitivity
to statistical fluctuations among the higher moments. For a sample
of $N$ quantiles, ``base'' bin counts were accumulated and stored
in the vector, $n(m)_{\epsilon}$, which represented counts in
$M_{max}$ bins.

\subsection{Simulating non-intermittent null distributions}
Non-intermittent null distributions were simulated once per intermittent distribution,
and the results were used to fill counts in $M_{max}$ total bins when the bin width was $\epsilon$.
First, the underlying smooth function of the histogram for
intermittent data was determined using kernel density estimation
(KDE) \cite{fadda} in the form
\be
f(m)=\frac{1}{Nh}\sum_{i=1}^N K \biggl( \frac{y_i - y_m}{h}
\biggr),
 \ee where $f(m)$ is the bin count for
the $m$th bin for non-intermittent null data, $N$ is the total
number of variates, $h=1.06 \sigma N^{-0.2}$ is the optimal
bandwidth for a Gaussian \cite{silverman}, and $\sigma$ is the
standard deviation of the Gaussian. $K$ is the Epanechnikov kernel
function \cite{epanech} defined as
\be
K(u)=
\begin{cases}
\frac{3}{4}(1-u^2) & |u|\leq1 \\ 0 & \textrm{otherwise,}
\end{cases}
\ee where $u=(y_i-y_m)/h$ and $y_m$ is the lower bound of the
$m$th bin.  The smooth pdf derived from KDE was used with the
rejection method to build-up bin counts, $n(m)_{null,\epsilon}$,
until the number of variates was equal to the number of variates
in the single-event intermittent distribution.  Under the
rejection method, bins in the simulated non-intermittent
distribution are randomly selected with the formula
\be
m=(M_{max}-1)\textrm{U}(0,1)_1 + 1. \ee  where $U(0,1)_1$ is a
pseudo-random uniform distributed variate.  For each $m$, a second
pseudo-random variate is obtained and if the following criterion
is met
\be
\textrm{U}(0,1)_2<
\textrm{pdf}(m)/\textrm{max}\{\textrm{pdf}(i)\}, \quad \quad
i=1,2,...,M_{max} \ee then one is added to the running sum for bin
count and the running sum for the total number of simulated $y$
values.  The rejection method provided non-intermittent null
distributions with attendant statistical fluctuations to determine
whether non-intermittent data consistently has estimates of
intermittency lower than that of a single-event distribution with
simulated intermittency.

\subsection{$F_2$ calculations and collapsing bin counts into $M$ Bins}
$F_2$ was calculated using (4) at varying values of $M=M_{max}/k$,
where $k$ is the number of bins at the experimental resolution
collapsed together ($k=2,3,...,M_{max}/30$). It follows that for
$M$ equally-spaced non-overlapping bins, the bin width is
$k\epsilon$. The smallest bin width of $2\epsilon$ ($M=M_{max}/2$)
used in $F_2$ calculations allowed us to conservatively avoid
artificial effects, whereas the greatest bin width was limited to
$\Delta y/30$ ($M=30$) since widths can become comparable to the
width of the distribution.  As $M$ changed, histogram bin counts
$n(m)$ were determined by collapsing together each set of $k$
contiguous bins at an assumed experimental resolution, i.e.,
$n(m)_{\epsilon}$ or $n(m)_{null,\epsilon}$, rather than
determining new lower and upper bin walls and adding up counts
that fell within the walls (Table 1). This approach cut down on a
tremendous amount of processor time while avoiding cumulative
rounding effects from repeatedly calculating new bin walls.

\subsection{Shifting the range of $y$}
A single ``shift'' was performed in which the full range $\Delta
y$ was moved by a value of $\epsilon=0.01$ followed by a repeat of
$F_2$ calculations, log-linear fits, and permutations. To
accomplish this, we varied the starting bin $n(shift)_{\epsilon}$
before collapsing bins.  For example, if $M_{max}=400$, the first
value of $M$ was $M_{max}=400/2=200$, since the first value of $k$
is 2. Bin counts for the 200 bins were based on collapsing
contiguous pairs $(k=2)$ of bins starting with $n(1)_{\epsilon}$
when $shift=1$. $F_2$ was then calculated, linear fits with
permutations were made, and values of $\ln(F_2)$ and $\ln(M)$ were
stored.  This was repeated with $shift=2$ so that pairs of base
bin were collapsed again but by starting with base bin
$n(2)_{\epsilon}$ when $shift=2$. The process of shifting the
range of $y$ and repeating the randomization tests for each
intermittent single-event distribution resulted in an entirely
different set of bin counts, increasing the variation in
intermittent and null data sets. Shifting the range and
re-calculating $F_2$ was performed in the original Bialas and
Peschanski paper (See Fig.\ 3, 1986).

\subsection{Permutation-based log-linear regression}
For intermittency and non-Gaussian distributions, it is unlikely
that the null distributions are thoroughly known. Therefore,
instead of using a single fit of the data to determine
significance for each coefficient, permutation-based log-linear
fits were used in which fit data were permuted (randomly
shuffled) and refit in order to compare results before and after
permutation.  Permutation-based regression methods are useful when
the null distributions of data used are unknown.

The presence of intermittency was determined by incorporating
values of $\ln(F_2)$ and $\ln(M)$ into a log-linear regression
model to obtain direct estimates for the difference in the
$y-$intercept and slope of $\ln(F_2)$ vs.\ $\ln(M)$ for
intermittent and null data. The model was in the form \be \ln(F_2)
=\beta_0 + \beta_1 \ln(M) + \beta_2 I(null) + \beta_3
\ln(M)I(null), \ee
 where $\beta_0$ is the
$y$-intercept of the fitted line $\ln(F_2)$ vs.\ $\ln(M)$ for
intermittent data, $\beta_1$ is the slope of the fitted line for
intermittent data, $\beta_2$ is the difference in $y$-intercepts
for intermittent and null, $I(null)$ is 1 if the record is for
null data and 0 otherwise, and $\beta_3$ represents the difference
in slopes for the intermittent and null data.  Results of modeling
suggest that $\beta_1$ can be significantly positive over a wide
variety of conditions.  When $\beta_3<0$, the slope of the
intermittent fitted line for $\ln(F_2)$ vs.\ $\ln(M)$ is greater
than the slope of $\ln(F_2)$ vs.\ $\ln(M)$ for null data, whereas
when $\beta_3>0$ the slope of intermittent is less than the slope
of null. Values of $\beta_0$ and $\beta_2$ were not of central
importance since they were used as nuisance parameters to prevent
forcing the fits through the origin.  While $\beta_1$ tracks with the degree of absolute
intermittency in the intermittent distribution, the focus is on
the difference in slopes between the intermittent single-event
distribution and the non-intermittent null distribution
characterized by $\beta_3$.  Again, $\beta_3$ is negative whenever
the slope of  $\ln(F_2)$ vs.\ $\ln(M)$ is greater in the
intermittent than the null. Strong negative values of $\beta_3$
suggest higher levels of intermittency on a comparative basis.

For each fit, the Wald statistic was first calculated as
$Z_j=\beta_j/\textrm{s.e.}(\beta_j)$.  Next, values of $I(null)$,
that is 0 and 1, were permuted within the fit data records and
the fit was repeated to determine $Z_j^{(b)}$ for the $b$th
permutation where $b=1,2,...,B$.  Permutations of $I(null)$
followed by log-linear fits were repeated 10 times $(B=10)$ for
each intermittent distribution. Since there were 10 permutations
per log-linear fit, 10 null distributions simulated via KDE and
the rejection method, and 2 shifts in $\Delta y$ per intermittent
distribution, the total number of permutations for each
intermittent distribution was $B=200$.  After $B$ total
iterations, the $p$-value, or statistical significance of each
coefficient was

\be
p_j=\frac{\# \{b: |Z_j^{(b)}| > |Z_j| \}}{B}
\end{equation}

If the absolute value of the Wald statistic $Z_j^{(b)}$ at any
time after permuting $I(null)$ and performing a fit exceeds the
absolute value of the Wald statistic derived before permuting
$I(null)$, then there a greater undesired chance that the
coefficient is more significant as a result of the permuted
configuration. What one hopes, however, is that fitting after
permuting labels never results in a Wald statistic that is more
significant that that based on non-permuted data.  Because test
statistics (e.g., Wald) are inversely proportional to variance, it
was important to use $Z_j$ as the criterion during each
permutation.

\subsection{Statistical significance of fit coefficients}

The basis for tests of statistical significance is established by
the null hypothesis.  The null hypothesis states that there is no
difference in intermittency among intermittent and null
distributions whereas the alternative hypothesis posits that the
there is indeed a difference in intermittency.  Formally stated, the null hypothesis is
$H_o: \beta_3=0$ and the one-sided alternative hypothesis is $H_a: \beta_3<0$, since negative values of
$\beta_3$ in (11) imply intermittency.
The goal is to discredit the null hypothesis.  A false positive test result
causing rejection of the null hypothesis when the null is in fact
true is known as a Type I error.  A false negative test result
causing failure to reject the null hypothesis when it is not true
(missed the effect) is a Type II error.  The probability of making
a Type I error is equal to $\alpha$ and the probability of making
a Type II error is equal to $\beta_{power}$.  Commonly used
acceptable error rates in statistical hypothesis testing are
$\alpha=0.05$ and $\beta_{power}=0.10$.  The Wald statistics
described above follow the standard normal distribution, such that
a $Z_j$ less than -1.645 or greater than 1.645 lies in the
rejection region where $p<0.05$.  However, since permutation-based
regression models were used, the significance of fit coefficients was not
based on comparing Wald statistics with standard normal rejection regions, but rather
the values of the empirical $p$-values in (12).  Whenever $p_j<\alpha$, a
coefficient is said to be significant and the risk of a Type I
error is at least $\alpha$.

\subsection{Statistical power}

Randomization tests were used for assessing the level of
significance for each fit coefficient, given the intermittent
single-event and the simulated null distributions used. When
developing a test of hypothesis, it is essential to know the power
of the test for each log-linear coefficient.  The power of a
statistical test is equal to $1-\beta_{power}$, or the probability
of rejecting the null hypothesis when it is not true.  In other
words, power is the probability of detecting a true effect when it
is truly present.  Power depends on the sample size $N$ of the
Gaussian, effect size $\beta_j$, and level of significance
($\alpha$). To determine power for each fit coefficient, 1000
intermittent single-event Gaussians with the same induced spikes
and holes and sample size were simulated, and used in 1000
randomizations tests.  During the 1000 randomization tests, power for each fit
coefficient was based on the proportion of tests in which $p_j$ in (12) was less than $\alpha=0.05$.
For a given effect size $\beta_j$, sample size of Gaussian $N$, and significance level
($\alpha$), an acceptable level of power is typically greater than
0.70.  Thus, the $p$-value for a particular log-linear fit
coefficient would have to be less than 0.05 during 700
randomization tests on an assumed intermittent distribution.

\subsection{Simulated intermittent distributions used}
Three types in intermittent distributions were simulated, each having a single hole-spike
pair at a different location.  Hole location and widths are discussed first.
The first simulated intermittent distribution had an induced
hole starting at $y=$0.2 of width $\Delta_h$=0.64, corresponding to the interval (0.2-0.84)$_h$.
The second distribution had a hole starting at $y=1.0$ of width $\Delta_h$=0.02, corresponding to
interval (1.0-1.2)$_h$.  The third type of intermittent distribution had a single hole
induced starting at $y=$2 also at a width of $\Delta_h$=0.64, corresponding to
interval (2.0-2.64)$_h$.  For each hole induced, a single spike was generated by adding hole data
to existing simulated data below the $y$-values using
varying spike widths of $\Delta_s$=0.64, 0.24, 0.08, or 0.02.
Therefore, for the first distribution with a hole in interval (0.2-0.84)$_h$, spikes were either in
intervals (-0.44-0.2)$_s$, (-0.04-0.2)$_s$, (0.12-0.2)$_s$, or (0.18-0.2)$_s$.  For the distribution with
a hole in interval (1.0-1.2)$_s$, the spike intervals were at either (0.36-1.0)$_s$, (0.76-1.0)$_s$, (0.92-1.0)$_s$,
or (0.98-1.0)$_s$.
Finally, for a hole in interval (2.0-2.64)$_h$, spike intervals were either at (1.36-2.0), (1.76-2.0),
(1.92-2.0), or (1.98-2.0).

The combination of distribution having single hole-spike pairs at three hole locations, and four spike
locations resulted in a total of 12 distributions.  Each of the 12 types of intermittent
distributions were simulated with 500, 1000, 1500, 2000, 4000, 8000, 16000, and 32000 variates.
This resulted in 96 different simulated intermittent distributions (Table 2).

\subsection{Algorithm flow}
The algorithm flow necessary for performing a single randomization test
to determine significance of fit coefficients for one single-event distribution is given in
Figures 4 and 5.  Figure 4 shows how test involves 10 simulations of
non-intermittent null distributions, 2 shifts each, and $B=10$ permutations
during the log-linear fit.  The 10 null distribution simulations
and 10 permutations are done twice, once for the first shift based on interval
$y_{min}$ to $y_{max}$ and once for the second shift using the interval
$y_{min}+\epsilon$ to $y_{max}+\epsilon$.

Figure 5 shows a telescoping schematic of what occurs for each of
the 10 null distributions.  For each null distribution there are
two shifts during which $\ln(F_2)$ and $\ln(M)$ are determined.
Values of $\ln(F_2)$ and $\ln(M)$ as a function of $M$ for
both the intermittent single event and null distribution are used
in the log-linear fit to yield $Z_j$ for each coefficient.  The
values of $I(null)$ are then permuted $B=10$ times resulting in
$Z_j^{(b)}$, where $b=1,2,...,10$, for a total of 200 permutations (10*2*10).
Bookkeeping is done using the counter \texttt{numsig}$_j$ which keeps track of the total number
of times $|Z_j^{(b)}|$ exceeds $|Z_j|$ during the 200 permuted
fits. The 200 fits represent one randomization test, from which
the statistical significance of each fit coefficient is determined
based on $p_j$ in (12).  Statistical power for each coefficient was determined
based on 1000 randomization tests in which the same intermittent
distribution (same hole and spike widths and sample size) was
simulated.

\subsection{Summary statistics of results}
Each randomization test to determine the statistical significance
of fit coefficients for a single simulated intermittent
distribution employed 200 permutation-based log-linear fits.  The
200 total fits per randomization test (i.e., intermittent
distribution) is based on 10 simulations to generate
non-intermittent null distributions, followed by 2 shifts, and
then 10 permutation-based fits (10*2*10).  Within each
randomization test, the total 200 permutation-based fits were used
for obtaining the significance of each coefficient via $p_j$ in
(12).  Among the 200 total permutation-based fits, there were 20
fits for which fit data were not permuted.  Averages of fit
coefficients and their standard error were obtained for the 20
``non-permuted'' fits.  Since randomization tests were carried out
1000 times (in order to obtain power for each coefficient), the
averages of coefficients and their standard errors over the 1000
tests was based on averages (over 1000 randomization tests) of
averages (over 20 non-permuted fits).  The global averages of coefficients and
their standard errors, based on a total of 20,000 non-permuted fits,
are listed in Table 2 under column headers with angle brackets
$\langle \rangle$.  In total, since there were 20 non-permuted and 200 permuted fits per
randomization test, the use of 1000 randomization tests for obtaining power for
each coefficient for the intermittent distribution considered resulted in 200,000
permutation-based fits for each distribution (row in Table 2).

\parskip 0pt
\parindent 10pt

\section{Results}
The averages of fit coefficients and their standard deviation from 20,000 non-permuted
fits during 1000 randomization tests are listed in
of Table 2.  Each row in Table represents the result of 1000 randomization tests for a single simulated
intermittent distribution.  Also
listed in the Table 2 is the statistical power for each
coefficient reflecting the proportion of the 1000 randomization tests during which
each coefficient was significant.

As described in the methods section, $\beta_0$ reflects the
$y$-intercept of the fitted line for $\ln(F_2)$ vs.\ $\ln(M)$ for
intermittent data. Average values of $\beta_0$ were jumpy and did
not correlate consistently with the level of induced
intermittency.  There can be tremendous variation
in the $y$-intercept of either an intermittent or null
distribution depending on unique properties of the distributions, so this was expected.

Average values of $\beta_1$ reflected the slope of fitted line
$\ln(F_2)$ vs.\ $\ln(M)$ for intermittent data, and ranged from
0.0007 to 0.2347 for distributions with 32000 variates. One of the
most interesting observations was that, as sample size of the
intermittent distributions decreased from 32000 to 500, values of
$\beta_1$ increased. This was expected because the
abundance of artificial holes in histograms increases as the sample
size of the input distribution decreases. Accordingly, the
``true'' intermittency effect induced is portrayed at only the
larger sample sizes where there is less likelihood for artificial
holes to exist in the histograms.

Similar to $\beta_0$, $\beta_2$ reflects the difference between
the $y$-intercepts of the intermittent and null distributions
compared.   But again, $\beta_0$ and $\beta_2$ are essentially
important nuisance parameters introduced in the model in order not
to force the fitted lines through the origin.  Nevertheless, $\beta_2$ was found
to be especially important when evaluating power as will be
discussed later.

The most important coefficient was $\beta_3$, which reflects the
difference in slopes between fitted lines $\ln(F_2)$ vs.\ $\ln(M)$
for intermittent and non-intermittent data.  (Values of $\beta_3$
and $P(\beta_3)$ when $N$=32000 are listed in bold in Table 2).
When $\beta_3<0$, the slope of the line for $\ln(F_2)$ vs.\
$\ln(M)$ of the intermittent data is greater than the slope of
$\ln(F_2)$ vs.\ $\ln(M)$ for non-intermittent data, whereas when
$\beta_3>0$, the slope of intermittent is less than the slope of
non-intermittent. Average values for $\beta_3$ for distributions
with 32000 variates ranged from -0.5294 to -0.0002.  The
statistical power $P(\beta_3)$ of $\beta_3$ for detecting the
induced intermittency in various intermittent distributions is
also shown in Table 2. For distributions with 32000 variates,
$P(\beta_3)$ was 100\% for values of $\beta_3$ that were more
negative than -0.015, suggesting that at large samples sizes the
randomization test had 100\% power to detect a
difference of 0.015 between slopes for intermittent and null
distributions.  It was observed that for these cases, $P(\beta_3)$
was 100\% only when the $y$-intercept difference $\beta_2$
exceeded 0.05.  In one case, $\beta_3$ was more negative than
-0.015 (-0.0319), but $P(\beta_3)$ was 82.8\%.  For
this case, $\beta_2$ was equal to 0.0385.  For the remaining
distributions for which $N$=32000, when $\beta_3$ was more
positive than -0.015, $P(\beta_3)$ was less than 100\%.

It was also noticed that while $\beta_1$ increased with decreasing
sample size as a result of the introduction of artificial holes,
$P(\beta_3)$ decreased.  This is important because it suggests
that the power of the randomization test to detect intermittency
decreases with decreasing sample size.

Power of the randomization test was also assessed for non-intermittent null distributions (end of Table 2).
The null distributions used did not contain any holes or spikes and therefore did not contain
induced intermittency.  Since intermittency is not present in null distributions, zero power is expected.
With regard to coefficient values for null distributions, $\beta_0$ did not differ from values of $\beta_0$ for intermittent
distributions, but $\beta_1$ increased from 0.0004 for $N$=32000 to 0.0389 for $N$=500.
As sample sized decreased from $N$=32000 to $N$=500, $\beta_2$ increased from -0.0016 to
0.1471 and $\beta_3$ decreased from 0.00005 to -0.0294, suggesting again that even in null
distributions the level of artificial intermittency increases as sample size decreased.

The value of the power calculations is reflected in results for null distributions lacking any effect,
because power indicates the probability that the $p-$values of coefficients in (12) are
significant (i.e., $p_j<\alpha=0.05$).  The greatest probability of obtaining a significant
value for either $\beta_2$ or $\beta_3$ for a single randomization test on a null
distribution was 0.062 for $N$=1000.  At $N=500$ the probability (power) of obtaining a significant coefficient
for $\beta_2$ or $\beta_3$ was 0.04 and 0.03, respectively.  At $N=2000$ the power for $\beta_2$ or $\beta_3$ was 0.015 and 0.019, respectively.
At $N=8000$ and greater the probability (power) of obtaining a significant coefficient
for $\beta_2$ or $\beta_3$ was zero.  In conclusion, power was zero for null
distributions having sample sizes greater than $N=8000$.

Table 3 lists the sorted values of statistical power $P(\beta_3)$ and
averages of coefficients extracted from Table 2 for distributions with N=32000.
The performance of $\beta_3$ for detecting the induced intemittency in terms of slope
difference between intermittent and null distributions can be gleaned from this table.
For most experiments, it is desired to use a test that has 80\% or more power to detect a
signal as a function of effect size, sample size and $\alpha$.  In this study, effect
size is reflected in the $\beta_j$ coefficients.   Sample size is reflected in the number
of variates $N$ used for each simulated distribution.  The significance level
$\alpha$ is used as an acceptable probability of Type I error, which essentially means reporting
that a coefficient is significant (rejecting the null hypothesis of no signal) when the coefficient is truly
insignificant (false positive).  One the other hand, power is one minus the Type II error rate, which is the
probability of missing a signal when the signal is present (false negative).  One minus this
probability (i.e., power) is the probability of finding a signal when it is truly
present.  Looking at Table 3, one notices that when $\beta_3$ was more negative than -0.01
(greater than a 1\% signal), the power of $\beta_3$ for detecting intermittency was at least 70\%.

Figure 6 shows the statistical power $P(\beta_3)$ as a function of
average effect size $\langle \beta_3 \rangle$ and sample size of simulated intermittent
Gaussian distributions.  Each point represents a distribution in a row of Table 2.  At
the lowest sample size of $N$=500, an acceptable level of approximately 70\% power is
attainable for a 0.08 difference in slopes between the intermittent and null
distributions.  As sample size increases, the power to detect intermittency increases.
For intermittent Gaussians with 4000 variates, there was approximately 70\% power to detect
a slope difference of 0.02 between intermittent and null distributions. Whereas for
sample sizes of 8000 and greater, there was more than 70\% power to detect a
slope difference of 0.01.

\section{Discussion}
A basic characteristic of QCD intermittency is that if the smooth
distribution for a histogram measured at the limit of experimental
resolution is discontinuous, it should reveal an abundance of
spikes and holes.  For a discontinuous smooth distribution,
QCD intermittency is a measure of non-normality in fluctuations
and reflects little about the deterministic or stochastic
properties of a distribution. QCD intermittency is also
independent of the scale of data and scaling in spatial or
temporal correlations. The lowest scale of resolution used in this
paper ($\delta y=0.01$) refers to a measure of imprecision, or
standard deviation in measured data. Thus, the quantile values of
Gaussians used were not assumed to be infinitely precise.

The statistically significant levels of intermittency identified
in this study show how various methods from applied statistics can
be assembled to form a randomization test for intermittency for
single-event distributions.  It is important to compare fit
coefficients for the intermittent distribution versus that from a
null distribution.  This study employed a log-linear model that
was able to extract simultaneously information on the
$y$-intercepts and slopes for intermittent and null data
separately.  At the same time, the log-linear model provided a
method to simultaneously test the statistical significance of each
coefficient.  Not surprisingly, the most important and consistent
coefficient for identifying intermittency was $\beta_3$, or the
difference in slopes of the line $\ln(F_2)$ vs.\ $\ln(M)$ for
intermittent and null distributions.

Generally speaking, the power to detect intermittency exceeds 70\%
when $\beta_3$ is less than (more negative than) -0.01 assuming a
Gaussian sample size of 32000.  As the sample size decreases,
artificial holes are introduced which obscures any real
intermittency.  This was reflected in
an increase of $\beta_1$ with decreasing sample size.
As the sample size decreased, wider artificial holes were introduced into the histograms for both
the intermittent and non-intermittent null distributions.  Since both distributions
were affected equally on a random basis, the difference in slopes measured
by $\beta_3$ did not increase.  Power naturally increases with sample size,
so there was no reason to expect an increase in power with decreasing sample size.
Nevertheless, a positive finding was that the power of the
randomization test to detect intermittency decreased with
decreasing sample size, in spite of the increase of $\beta_1$.
This indicates that the ability to appropriately identify the
presence of intermittency in a single-event distribution depends
on more than a single linear fit of $\ln(F_2)$ vs.\  $\ln(M)$ for
the intermittent distribution.

The finding that power was at a maximum of 0.06 for $\beta_3$ from a null distribution
of size 1000 suggests that there is at most a 6\% chance of obtaining a $p$-value for $\beta_3$ below 0.05.
For null distributions of sample size 4000, the probability (power) of obtaining a significant $\beta_3$
coefficient was 0.007.  When sample size was 8000 or more, power for $\beta_3$ was zero.
No attempt was made to relate hole and spike widths and their locations in simulated intermittent Gaussians
with resultant power, since the effect size needed for power calculations
was captured by the fit coefficients.  This was confirmed by the change in coefficients
with changing sample size.  While the widths and locations of induced holes and spikes were
fixed, coefficients changed with sample size and the induced effect.  The purpose of
power is to establish the consistency of a coefficient for detecting an effect when it is
truly present.  It follows that null distributions are also not used for establishing
power of a test, since power depends on a known effect size $\beta_j$, sample size $N$ of a Gaussian,
and the level of significance $\alpha$ used for determining when each coefficient is
significant.

There were several limitations encountered in this study. First,
there is an infinitely large number of ways one can induce
intermittency in Gaussian distributions.  Given the size and
duration of the study, it was assumed that the 96 intermittent
distributions considered would adequately reflect the robustness
of the randomization test over a range of induced levels of intermittency.
Additional research is needed to assess the power of the
randomization test for horizontal, vertical, and mixed SFMs,
number of null distributions, number of shifts, number of permutations during fits,
distribution sample sizes, imprecision, skewness and kurtosis of multimodal
distributions, and variations in the choice of bandwidth and
kernel smoother, etc.  It was impossible to address the majority
of these issues in this investigation, and therefore the intent of
this paper was to introduce results for the limited set of
conditions used.

Over the course of this investigation, there were many evaluations
on how best to make a smooth distribution for the non-intermittent
null distributions.  The most promising was
the combined approach using KDE and the rejection method, which is also
probably the most robust.  Parametric methods were
problematic when there were large holes or spikes, and when there was a
significant level of kurtosis or skewness in the data.  Long tails
present another challenge for simulating an appropriate null
distribution with parametric fitting methods.  KDE is
non-parametric and by altering the bandwidth settings one can
closely obtain the original histogram, or more smoothed
histograms.  Because the simulated distributions were known be
standard normal Gaussian distributions, the optimal bandwith
$h=1.06 \sigma N^{-1/5}$ was used \cite{silverman}.

If an experimenter has a single-event distribution for which
intermittency is in question, then application of sequential
methods listed in steps 2-7 skipping simulation in step 1 would be
used.  This corresponds to running a single randomization test on
the intermittent data.  If the $p$-value in (12) for any of the
coefficients in (11) are less than 0.05, then the coefficient is
statistically significant.  Specifically, intermittency would be
statistically significant if the $p$-value for $\beta_3$ was less
than 0.05.  Power of the randomization test for truly detecting
intermittency could be looked up in Figure 6, for the specific
coefficient value of $\beta_3$ and sample size of the
distribution.  An unacceptable value of power below 0.70 would
suggest that a greater effect size or greater sample size is
needed for detecting a statistically significant level of
intermittency.

\section{Conclusions}
Results indicate that the slope of $\ln(F_2)$ vs.\ $\ln(M)$ for
intermittent distributions increased with decreasing sample size,
due to artificially-induced holes occurring in sparse histograms.

When the average difference in slopes between intermittent
distributions and null distributions was greater than 0.01, there
was at least 70\% power for detecting the effect for
distributions of size 8000 and greater.  The randomization test performed
satisfactorily since the power of the test for intermittency
decreased with decreasing sample size.

\section{Acknowledgments}

The author acknowledges the support of grants CA-78199-04/05, and CA-100829-01 from
the National Cancer Institute, and helpful suggestions from K.\ Lau.

\clearpage

\begin{table}
\begin{center}
\caption{Example of collapsing bins to calculate bin counts
$n(m)$, as the total number of histogram bins ($M$) change with
each change of scale $\delta y$. $k$ is the number of bins added
together to obtain bin counts for each of $M$ total bins}.
\begin{tabular}{|c|c|c|l|l|l|l|l|l|l|l|l|l|l|} \hline $\delta y$&
$k$& $M$ & \multicolumn{11}{|c|}{Bin counts, $n(m)$}
\\\hline 0.06& 6& $M_{max}/6$  & & & & & & 23& & & & & 68
\\ \hline 0.05& 5& $M_{max}/5$ & & & & & 14& & & & & 60&
 \\
\hline 0.04& 4& $M_{max}/4$ & & & & 7& & & & 44& & &
 \\
\hline 0.03& 3& $M_{max}/3$ & & & 3& & & 20& & & 40& &
 \\
\hline 0.02& 2& $M_{max}/2$ & & 1& & 6& & 16& & 28& & 23&
 \\
\hline 0.01=$\epsilon*$  &  1 & $M_{max}$ & 1& 0& 2& 4& 7& 9& 13&
15& 12& 11& 8 \\ \hline
\end{tabular}
\begin{flushleft}\parindent -7pt
 \hspace{1in} * $\epsilon $ is the assumed experimental
imprecision (standard deviation) equal to 0.01.
\end{flushleft}
\label{tab1}
\end{center}
\end{table}

\small

\begin{sidewaystable}
\begin{center}
\caption{Averages of log-linear fit coefficients and their
standard errors and statistical power for each coefficient.
Averages based on 20,000 fits and power based on 1000 tests.}
\begin{tabular}{|c|c|c|c|c|c|c|c|c|c|c|c|c|c|}
\hline Interval$^a$& $N^b$ & $\langle\beta_0\rangle^c$&
$\langle\sigma_{\beta_0}\rangle$& $\langle\beta_1\rangle$&
$\langle\sigma_{\beta_1}\rangle$& $\langle\beta_2\rangle$&
$\langle\sigma_{\beta_2}\rangle$& $\langle\beta_3\rangle$&
$\langle\sigma_{\beta_3}\rangle$ & P($\beta_0$)$^d$& P($\beta_1$)&
P($\beta_2$) & P($\beta_3$)
\\ \hline (0.2-0.84)$_h$& 500& 0.6668& 0.1675& 0.0698& 0.0219& 0.2654&
0.1584& -0.0793& 0.0337& 0.700& 0.939& 0.495& 0.606 \\ \hline
(-0.44-0.2)$_s$& 1000& 0.8117& 0.1867& 0.0510& 0.0273& 0.2178&
0.2083& -0.0662& 0.0438& 0.901& 0.924& 0.506& 0.661 \\ \hline &
1500& 0.9020& 0.1851& 0.0389& 0.0274& 0.1417& 0.2094& -0.0484&
0.0440& 0.971& 0.951& 0.373& 0.635 \\ \hline & 2000& 0.9642&
0.1724& 0.0318& 0.0248& 0.0955& 0.1914& -0.0375& 0.0402& 0.987&
0.976& 0.260& 0.676 \\ \hline & 4000& 1.0757& 0.0968& 0.0191&
0.0117& 0.0083& 0.0867& -0.0163& 0.0180& 1.000& 0.996& 0.043&
0.761 \\ \hline & 8000& 1.1318& 0.0593& 0.0169& 0.0041& 0.0002&
0.0223& -0.0121& 0.0044& 1.000& 0.999& 0.011& 0.810 \\ \hline &
16000& 1.1723& 0.0507& 0.0176& 0.0036& 0.0087& 0.0170& -0.0118&
0.0032& 1.000& 1.000& 0.046& 0.792 \\ \hline & 32000& 1.2085&
0.0475& 0.0181& 0.0036& 0.0150& 0.0176& \textbf{-0.0113}& 0.0033&
1.000& 1.000& 0.099& \textbf{0.763} \\ \hline (0.2-0.84)$_h$& 500&
0.7117& 0.1858& 0.1062& 0.0267& 0.2630& 0.1936& -0.1218& 0.0418&
1.000& 0.998& 0.830& 0.893 \\ \hline (-0.04-0.2)$_s$& 1000&
0.8872& 0.2136& 0.0821& 0.0322& 0.1813& 0.2473& -0.0969& 0.0522&
1.000& 0.999& 0.565& 0.922 \\ \hline & 1500& 0.9654& 0.2046&
0.0715& 0.0315& 0.1242& 0.2433& -0.0802& 0.0513& 1.000& 1.000&
0.411& 0.920 \\ \hline & 2000& 1.0315& 0.1821& 0.0624& 0.0280&
0.0677& 0.2166& -0.0647& 0.0454& 1.000& 1.000& 0.264& 0.919 \\
\hline & 4000& 1.1476& 0.1111& 0.0488& 0.0161& -0.0129& 0.1129&
-0.0403& 0.0233& 1.000& 1.000& 0.051& 0.903 \\ \hline & 8000&
1.1910& 0.0700& 0.0488& 0.0100& -0.0004& 0.0521& -0.0356& 0.0099&
1.000& 1.000& 0.027& 0.899 \\ \hline & 16000& 1.2280& 0.0621&
0.0505& 0.0095& 0.0228& 0.0505& -0.0340& 0.0094& 1.000& 1.000&
0.087& 0.883 \\ \hline & 32000& 1.2571& 0.0654& 0.0519& 0.0098&
0.0385& 0.0520& \textbf{-0.0319}& 0.0097& 1.000& 1.000& 0.146&
\textbf{0.828} \\ \hline (0.2-0.84)$_h$& 500& 0.5614& 0.1444&
0.2141& 0.0295& 0.2647& 0.1263& -0.2058& 0.0316& 0.814& 1.000&
0.542& 0.995 \\ \hline (0.12-0.2)$_s$& 1000& 0.6253& 0.1759&
0.2101& 0.0345& 0.3105& 0.1670& -0.2055& 0.0348& 0.757& 1.000&
0.556& 0.999 \\ \hline & 1500& 0.6645& 0.2046& 0.2090& 0.0399&
0.3331& 0.2040& -0.2038& 0.0401& 0.623& 1.000& 0.445& 0.999 \\
\hline & 2000& 0.6588& 0.2176& 0.2124& 0.0419& 0.3599& 0.2161&
-0.2033& 0.0416& 0.533& 1.000& 0.407& 0.999 \\ \hline & 4000&
0.6761& 0.2264& 0.2173& 0.0430& 0.4184& 0.2277& -0.2036& 0.0427&
0.373& 1.000& 0.377& 0.997 \\ \hline & 8000& 0.6746& 0.2062&
0.2250& 0.0396& 0.4855& 0.2139& -0.2051& 0.0398& 0.260& 1.000&
0.460& 0.998 \\ \hline & 16000& 0.6908& 0.2023& 0.2289& 0.0384&
0.5225& 0.2070& -0.2005& 0.0383& 0.238& 1.000& 0.548& 1.000 \\
\hline & 32000& 0.6795& 0.2215& 0.2367& 0.0416& 0.5684& 0.2108&
\textbf{-0.1974}& 0.0386& 0.180& 1.000& 0.663& \textbf{1.000}
\\\hline
\end{tabular}
 \begin{flushleft} \footnotesize
 $^a$ Interval of artificial hole $( )_h$ and spike $( )_s$ in simulated intermittent Gaussians.  Hole data were added to existing data in spike interval.\\
 $^b$ Number of standard normal variates in each simulated Gaussian (sample size).\\
 $^c$ Averages in angle brackets $\langle \rangle$ based on total of 20,000 non-permuted
 fits (20 non-permuted fits per randomization test times 1000 tests).\\
 $^d$ Statistical power for each fit coefficient based on proportion times each coefficient was significant among the 1000 tests.  \\
 \hspace{.1in} Power calculations employed 220,000 permutation-based fits (20 non-permuted fits and 200 fits with permutation per test times 1000 tests).  \\
 \end{flushleft}

\end{center}
\end{sidewaystable}

\small
\setcounter{table}{1}

\begin{sidewaystable}
\begin{center}
\caption{(cont'd).}
\begin{tabular}{|c|c|c|c|c|c|c|c|c|c|c|c|c|c|}
\hline Interval$^a$& $N^b$ & $\langle\beta_0\rangle^c$&
$\langle\sigma_{\beta_0}\rangle$& $\langle\beta_1\rangle$&
$\langle\sigma_{\beta_1}\rangle$& $\langle\beta_2\rangle$&
$\langle\sigma_{\beta_2}\rangle$& $\langle\beta_3\rangle$&
$\langle\sigma_{\beta_3}\rangle$ & P($\beta_0$)$^d$& P($\beta_1$)&
P($\beta_2$) & P($\beta_3$)
\\
 \hline (0.2-0.84)$_h$ & 500& -0.0340&
0.3669& 0.4017& 0.0838& 0.2730& 0.6358& -0.2756& 0.1302& 0.000&
1.000& 0.174& 1.000 \\ \hline (0.18-0.2)$_s$ & 1000& -0.2038&
0.4992& 0.4452& 0.1110& 0.7156& 0.8545& -0.3561& 0.1739& 0.000&
1.000& 0.401& 1.000 \\ \hline & 1500& -0.3956& 0.5029& 0.4900&
0.1110& 1.0951& 0.8552& -0.4278& 0.1749& 0.001& 1.000& 0.612&
1.000 \\ \hline & 2000& -0.5100& 0.4780& 0.5153& 0.1041& 1.3164&
0.7992& -0.4676& 0.1623& 0.000& 1.000& 0.725& 1.000 \\ \hline &
4000& -0.7414& 0.2587& 0.5693& 0.0554& 1.7832& 0.3768& -0.5496&
0.0770& 0.000& 1.000& 0.964& 1.000 \\ \hline & 8000& -0.7410&
0.1711& 0.5729& 0.0351& 1.8679& 0.1845& -0.5518& 0.0377& 0.000&
1.000& 0.998& 1.000 \\ \hline & 16000& -0.7276& 0.1581& 0.5743&
0.0318& 1.9082& 0.1589& -0.5442& 0.0324& 0.000& 1.000& 1.000&
1.000 \\ \hline & 32000& -0.6976& 0.1617& 0.5721& 0.0326& 1.9112&
0.1547& \textbf{-0.5294}& 0.0321& 0.000& 1.000& 1.000&
\textbf{1.000} \\ \hline (1-1.2)$_h$& 500& 0.3973& 0.1247& 0.0447&
0.0142& 0.1609& 0.0879& -0.0377& 0.0186& 0.207& 0.754& 0.077&
0.097 \\ \hline (0.36-1)$_s$& 1000& 0.5503& 0.1461& 0.0262&
0.0173& 0.1374& 0.1272& -0.0331& 0.0268& 0.459& 0.507& 0.091&
0.134 \\ \hline & 1500& 0.6061& 0.1407& 0.0198& 0.0172& 0.1040&
0.1297& -0.0261& 0.0273& 0.552& 0.489& 0.059& 0.137 \\ \hline &
2000& 0.6564& 0.1299& 0.0147& 0.0159& 0.0698& 0.1191& -0.0188&
0.0250& 0.664& 0.514& 0.045& 0.132 \\ \hline & 4000& 0.7505&
0.0783& 0.0061& 0.0069& 0.0086& 0.0508& -0.0057& 0.0106& 0.832&
0.643& 0.003& 0.144 \\ \hline & 8000& 0.7997& 0.0559& 0.0051&
0.0023& 0.0015& 0.0108& -0.0037& 0.0022& 0.921& 0.793& 0.000&
0.220 \\ \hline & 16000& 0.8456& 0.0507& 0.0051& 0.0017& 0.0027&
0.0051& -0.0035& 0.0009& 0.972& 0.885& 0.001& 0.270 \\ \hline &
32000& 0.8848& 0.0475& 0.0052& 0.0016& 0.0039& 0.0046&
\textbf{-0.0033}& 0.0008& 0.983& 0.917& 0.001& \textbf{0.248} \\
\hline (1-1.2)$_h$& 500& 0.3856& 0.1254& 0.0463& 0.0143& 0.1537&
0.0878& -0.0397& 0.0188& 0.233& 0.793& 0.084& 0.163 \\ \hline
(0.76-1)$_s$& 1000& 0.5216& 0.1394& 0.0301& 0.0166& 0.1429&
0.1223& -0.0377& 0.0259& 0.542& 0.665& 0.178& 0.284 \\ \hline &
1500& 0.5956& 0.1376& 0.0223& 0.0170& 0.0970& 0.1272& -0.0280&
0.0267& 0.734& 0.695& 0.166& 0.329 \\ \hline & 2000& 0.6477&
0.1206& 0.0167& 0.0150& 0.0597& 0.1126& -0.0200& 0.0236& 0.836&
0.742& 0.104& 0.324 \\ \hline & 4000& 0.7302& 0.0757& 0.0096&
0.0068& 0.0090& 0.0486& -0.0087& 0.0102& 0.980& 0.888& 0.023&
0.485 \\ \hline & 8000& 0.7844& 0.0543& 0.0085& 0.0026& 0.0040&
0.0136& -0.0067& 0.0027& 0.998& 0.968& 0.015& 0.648 \\ \hline &
16000& 0.8261& 0.0521& 0.0088& 0.0022& 0.0061& 0.0093& -0.0063&
0.0018& 1.000& 0.991& 0.022& 0.650 \\ \hline & 32000& 0.8624&
0.0481& 0.0091& 0.0021& 0.0086& 0.0086& \textbf{-0.0060}& 0.0016&
1.000& 0.995& 0.038& \textbf{0.664} \\ \hline (1-1.2)$_h$& 500&
0.3518& 0.1040& 0.0588& 0.0130& 0.1571& 0.0618& -0.0473& 0.0145&
0.437& 0.970& 0.270& 0.492 \\ \hline (0.92-1)$_s$& 1000& 0.4804&
0.1186& 0.0437& 0.0144& 0.1574& 0.0903& -0.0475& 0.0197& 0.928&
0.976& 0.604& 0.786 \\ \hline & 1500& 0.5425& 0.1174& 0.0379&
0.0136& 0.1297& 0.0916& -0.0416& 0.0196& 0.985& 0.991& 0.531&
0.894 \\ \hline & 2000& 0.5827& 0.1110& 0.0347& 0.0130& 0.1087&
0.0856& -0.0370& 0.0185& 0.998& 0.995& 0.507& 0.914 \\ \hline &
4000& 0.6531& 0.0762& 0.0300& 0.0090& 0.0754& 0.0519& -0.0292&
0.0110& 1.000& 1.000& 0.482& 0.982 \\ \hline & 8000& 0.7001&
0.0598& 0.0299& 0.0064& 0.0743& 0.0310& -0.0280& 0.0062& 1.000&
1.000& 0.607& 0.994 \\ \hline & 16000& 0.7402& 0.0550& 0.0304&
0.0056& 0.0787& 0.0275& -0.0277& 0.0053& 1.000& 1.000& 0.709&
1.000 \\ \hline & 32000& 0.7741& 0.0499& 0.0312& 0.0050& 0.0821&
0.0252& \textbf{-0.0270}& 0.0048& 1.000& 1.000& 0.785&
\textbf{1.000}
\\\hline
\end{tabular}
\end{center}
\end{sidewaystable}

\small
\setcounter{table}{1}

\begin{sidewaystable}
\begin{center}
\caption{(cont'd).}
\begin{tabular}{|c|c|c|c|c|c|c|c|c|c|c|c|c|c|}
\hline Interval$^a$& $N^b$ & $\langle\beta_0\rangle^c$&
$\langle\sigma_{\beta_0}\rangle$& $\langle\beta_1\rangle$&
$\langle\sigma_{\beta_1}\rangle$& $\langle\beta_2\rangle$&
$\langle\sigma_{\beta_2}\rangle$& $\langle\beta_3\rangle$&
$\langle\sigma_{\beta_3}\rangle$ & P($\beta_0$)$^d$& P($\beta_1$)&
P($\beta_2$) & P($\beta_3$)
\\
 \hline (1-1.2)$_h$& 500&
0.2511& 0.0894& 0.0865& 0.0217& 0.1590& 0.0947& -0.0541& 0.0212&
0.242& 1.000& 0.272& 0.713 \\ \hline (0.98-1)$_s$& 1000& 0.3316&
0.0846& 0.0813& 0.0200& 0.2345& 0.0966& -0.0705& 0.0220& 0.706&
1.000& 0.878& 1.000 \\ \hline & 1500& 0.3595& 0.0825& 0.0821&
0.0190& 0.2609& 0.0908& -0.0756& 0.0208& 0.779& 1.000& 0.975&
1.000 \\ \hline & 2000& 0.3709& 0.0867& 0.0842& 0.0180& 0.2827&
0.0860& -0.0799& 0.0196& 0.764& 1.000& 0.987& 1.000 \\ \hline &
4000& 0.4043& 0.0810& 0.0863& 0.0139& 0.3095& 0.0605& -0.0847&
0.0143& 0.825& 1.000& 1.000& 1.000 \\ \hline & 8000& 0.4559&
0.0765& 0.0858& 0.0105& 0.3140& 0.0457& -0.0842& 0.0105& 0.935&
1.000& 1.000& 1.000 \\ \hline & 16000& 0.4990& 0.0654& 0.0851&
0.0087& 0.3140& 0.0389& -0.0827& 0.0087& 0.980& 1.000& 1.000&
1.000 \\ \hline & 32000& 0.5362& 0.0616& 0.0843& 0.0079& 0.3112&
0.0360& \textbf{-0.0807}& 0.0078& 0.993& 1.000& 1.000&
\textbf{1.000} \\ \hline (2-2.64)$_h$& 500& 0.3690& 0.1315&
0.0366& 0.0140& 0.1488& 0.0838& -0.0296& 0.0182& 0.117& 0.613&
0.029& 0.027 \\ \hline (1.36-2)$_s$& 1000& 0.5061& 0.1396& 0.0217&
0.0162& 0.1431& 0.1170& -0.0298& 0.0248& 0.303& 0.272& 0.058&
0.055 \\ \hline & 1500& 0.5742& 0.1380& 0.0149& 0.0166& 0.1019&
0.1219& -0.0214& 0.0258& 0.419& 0.207& 0.032& 0.042 \\ \hline &
2000& 0.6220& 0.1269& 0.0100& 0.0149& 0.0683& 0.1132& -0.0145&
0.0237& 0.448& 0.166& 0.011& 0.016 \\ \hline & 4000& 0.7129&
0.0792& 0.0019& 0.0070& 0.0090& 0.0512& -0.0023& 0.0107& 0.384&
0.120& 0.000& 0.005 \\ \hline & 8000& 0.7691& 0.0565& 0.0007&
0.0017& -0.0015& 0.0049& -0.0001& 0.0010& 0.264& 0.120& 0.000&
0.001 \\ \hline & 16000& 0.8131& 0.0523& 0.0007& 0.0015& -0.0012&
0.0027& -0.0001& 0.0005& 0.192& 0.132& 0.000& 0.000 \\ \hline &
32000& 0.8532& 0.0483& 0.0007& 0.0014& -0.0008& 0.0017&
\textbf{-0.0002}& 0.0003& 0.150& 0.136& 0.000& \textbf{0.000} \\
\hline (2-2.64)$_h$& 500& 0.3681& 0.1320& 0.0375& 0.0139& 0.1484&
0.0838& -0.0298& 0.0183& 0.131& 0.632& 0.037& 0.037 \\ \hline
(1.76-2)$_s$& 1000& 0.5099& 0.1401& 0.0212& 0.0160& 0.1358&
0.1169& -0.0287& 0.0247& 0.329& 0.288& 0.051& 0.059 \\ \hline &
1500& 0.5696& 0.1350& 0.0151& 0.0163& 0.1010& 0.1225& -0.0216&
0.0257& 0.387& 0.214& 0.024& 0.034 \\ \hline & 2000& 0.6203&
0.1262& 0.0102& 0.0149& 0.0655& 0.1113& -0.0144& 0.0234& 0.489&
0.197& 0.010& 0.027 \\ \hline & 4000& 0.7107& 0.0780& 0.0024&
0.0070& 0.0090& 0.0518& -0.0026& 0.0108& 0.430& 0.155& 0.001&
0.007 \\ \hline & 8000& 0.7621& 0.0562& 0.0012& 0.0020& -0.0009&
0.0095& -0.0006& 0.0020& 0.319& 0.173& 0.000& 0.000 \\ \hline &
16000& 0.8057& 0.0500& 0.0012& 0.0015& -0.0008& 0.0030& -0.0006&
0.0005& 0.265& 0.201& 0.000& 0.000 \\ \hline & 32000& 0.8463&
0.0478& 0.0013& 0.0014& -0.0003& 0.0019& \textbf{-0.0005}& 0.0003&
0.188& 0.198& 0.000& \textbf{0.000} \\ \hline (2-2.64)$_h$& 500&
0.3563& 0.1282& 0.0398& 0.0130& 0.1506& 0.0768& -0.0314& 0.0168&
0.127& 0.707& 0.043& 0.040 \\ \hline (1.92-2)$_s$& 1000& 0.5059&
0.1385& 0.0235& 0.0155& 0.1365& 0.1121& -0.0302& 0.0236& 0.396&
0.449& 0.103& 0.121 \\ \hline & 1500& 0.5677& 0.1345& 0.0177&
0.0155& 0.1049& 0.1155& -0.0238& 0.0244& 0.471& 0.435& 0.072&
0.120 \\ \hline & 2000& 0.6161& 0.1225& 0.0128& 0.0140& 0.0705&
0.1047& -0.0167& 0.0220& 0.559& 0.465& 0.053& 0.129 \\ \hline &
4000& 0.7002& 0.0783& 0.0061& 0.0066& 0.0192& 0.0473& -0.0061&
0.0099& 0.701& 0.645& 0.035& 0.213 \\ \hline & 8000& 0.7521&
0.0569& 0.0048& 0.0019& 0.0100& 0.0067& -0.0041& 0.0014& 0.731&
0.743& 0.018& 0.304 \\ \hline & 16000& 0.7927& 0.0509& 0.0049&
0.0017& 0.0111& 0.0050& -0.0042& 0.0010& 0.833& 0.861& 0.032&
0.443 \\ \hline & 32000& 0.8373& 0.0498& 0.0050& 0.0016& 0.0120&
0.0041& \textbf{-0.0041}& 0.0008& 0.897& 0.903& 0.071&
\textbf{0.533}
\\\hline
\end{tabular}
\end{center}
\end{sidewaystable}

\small
\setcounter{table}{1}

\begin{sidewaystable}
\begin{center}
\caption{(cont'd).}
\begin{tabular}{|c|c|c|c|c|c|c|c|c|c|c|c|c|c|}
\hline Interval$^a$& $N^b$ & $\langle\beta_0\rangle^c$&
$\langle\sigma_{\beta_0}\rangle$& $\langle\beta_1\rangle$&
$\langle\sigma_{\beta_1}\rangle$& $\langle\beta_2\rangle$&
$\langle\sigma_{\beta_2}\rangle$& $\langle\beta_3\rangle$&
$\langle\sigma_{\beta_3}\rangle$ & P($\beta_0$)$^d$& P($\beta_1$)&
P($\beta_2$) & P($\beta_3$)
\\
 \hline (2-2.64)$_h$& 500&
0.3408& 0.1183& 0.0452& 0.0128& 0.1492& 0.0631& -0.0327& 0.0144&
0.210& 0.838& 0.117& 0.127 \\ \hline (1.98-2)$_s$& 1000& 0.4720&
0.1233& 0.0314& 0.0136& 0.1518& 0.0854& -0.0350& 0.0186& 0.720&
0.846& 0.450& 0.513 \\ \hline & 1500& 0.5277& 0.1203& 0.0271&
0.0132& 0.1317& 0.0892& -0.0311& 0.0191& 0.852& 0.912& 0.561&
0.682 \\ \hline & 2000& 0.5689& 0.1113& 0.0231& 0.0122& 0.1074&
0.0831& -0.0260& 0.0178& 0.939& 0.961& 0.634& 0.792 \\ \hline &
4000& 0.6468& 0.0777& 0.0177& 0.0067& 0.0672& 0.0406& -0.0176&
0.0088& 0.994& 0.999& 0.893& 0.992 \\ \hline & 8000& 0.7027&
0.0596& 0.0165& 0.0037& 0.0597& 0.0155& -0.0158& 0.0036& 1.000&
1.000& 0.996& 1.000 \\ \hline & 16000& 0.7473& 0.0545& 0.0161&
0.0025& 0.0584& 0.0086& -0.0154& 0.0021& 1.000& 1.000& 1.000&
1.000 \\ \hline & 32000& 0.7888& 0.0508& 0.0159& 0.0021& 0.0580&
0.0071& \textbf{-0.0150}& 0.0017& 1.000& 1.000& 1.000&
\textbf{1.000} \\
\hline
Null&
500&
0.3584&
0.1195&
0.0389&
0.0137&
0.1471&
0.0809&
-0.02935&
0.0174&
0.120&
0.694&
0.040&
0.030 \\
\hline
&
1000&
0.4967&
0.1332&
0.0219&
0.0159&
0.1387&
0.1152&
-0.02867&
0.0243&
0.320&
0.310&
0.054&
0.062 \\
\hline
&
1500&
0.5694&
0.1365&
0.0143&
0.0163&
0.0965&
0.1205&
-0.02010&
0.0253&
0.452&
0.216&
0.031&
0.038 \\
\hline
&
2000&
0.6239&
0.1197&
0.0080&
0.0138&
0.0533&
0.1042&
-0.01126&
0.0218&
0.504&
0.155&
0.015&
0.019 \\
\hline
&
4000&
0.7026&
0.0787&
0.0018&
0.0069&
0.0084&
0.0502&
-0.00200&
0.0105&
0.431&
0.098&
0.001&
0.007 \\
\hline
&
8000&
0.7591&
0.0571&
0.0003&
0.0016&
-0.0019&
0.0048&
0.00009&
0.0009&
0.261&
0.085&
0.000&
0.000 \\
\hline
&
16000&
0.8033&
0.0511&
0.0004&
0.0015&
-0.0016&
0.0028&
0.00003&
0.0005&
0.194&
0.120&
0.000&
0.000 \\
\hline
&
32000&
0.8434&
0.0496&
0.0004&
0.0014&
-0.0016&
0.0018&
\textbf{0.00005}&
0.0002&
0.141&
0.103&
0.000&
\textbf{0.000} \\
\hline
\end{tabular}
\end{center}
\end{sidewaystable}

\clearpage \large

\begin{table}[htbp]
\caption{Sorted values of statistical power $P(\beta_3)$ and averages of coefficients
from Table 2 for distributions with N=32000.}
\begin{center}
\begin{tabular}{|l|l|l|l|l|}
\hline $P(\beta_3)$ & $\langle \beta_0 \rangle $&  $\langle
\beta_1 \rangle $&  $\langle \beta_2 \rangle $&  $\langle \beta_3
\rangle $
\\ \hline 0.000& 0.8463& 0.0013& -0.0003& -0.0005 \\ \hline 0.000&
0.8532& 0.0007& -0.0008& -0.0002 \\ \hline 0.248& 0.8848& 0.0052&
0.0039& -0.0033
\\ \hline 0.533& 0.8373& 0.0050& 0.0120& -0.0041 \\ \hline 0.664&
0.8624& 0.0091& 0.0086& -0.0060 \\ \hline 0.763& 1.2085& 0.0181&
0.0150& -0.0113 \\ \hline 0.828& 1.2571& 0.0519& 0.0385& -0.0319
\\ \hline 1.000& -0.6976& 0.5721& 1.9112& -0.5294 \\ \hline 1.000&
0.6795& 0.2367& 0.5684& -0.1974 \\ \hline 1.000& 0.5362& 0.0843&
0.3112& -0.0807 \\ \hline 1.000& 0.7741& 0.0312& 0.0821& -0.0270
\\ \hline 1.000& 0.7888& 0.0159& 0.0580& -0.0150 \\ \hline
\end{tabular}
\label{tab1}
\end{center}
\end{table}

\clearpage
\parskip 15pt

Figure 1: Artificially-induced hole and spike in frequency
distribution caused by moving data between $y$ and $y+ \Delta_h$
and adding it to existing data between $y$ and $y- \Delta_s$.
$\Delta_h$ is the width of the hole and $\Delta_s$ is the width of
the spike formed by adding hole data.

Figure 2. Scaled factorial moments ($F_2$) resulting from an
artificially induced hole of width $\Delta_h$ beginning at $y$=2.0
in a frequency histogram of 10,000 standard normal variates. (a)
Bin counts $n(m)$ when bin width $\delta y=\epsilon=0.01$,
$\Delta_s$ is the interval width into which bin counts from the
hole were randomly distributed during replacement.  (b)
Characteristics of scaled factorial moments ($F_2$) as a function
of total bins $(M)$ based on varying $\delta y$.

Figure 3. Scaled factorial moments ($F_2$) resulting from an
artificially induced hole of width $\Delta_h$ beginning at $y$=0.2
in a frequency histogram of 10,000 standard normal variates. (a)
Bin counts $n(m)$ when bin width $\delta y=\epsilon=0.01$,
$\Delta_s$ is the interval width into which bin counts from the
hole were randomly distributed during replacement. (b)
Characteristics of scaled factorial moments ($F_2$) as a function
of total bins $(M)$ based on varying $\delta y$.

Figure 4. Algorithm flow for a single randomization test.  The
result of a randomization test is the number of times (i.e.,
\texttt{numsig}$_j$) $|Z_j^{(b)}|>|Z_j|$ for each coefficient.

Figure 5. Detailed schematic showing complete methodology for a
single randomization test.  Each test includes analysis for each
of the 10 null distributions.

Figure 6. Statistical power $P(\beta_3)$ to detect a significant
fit coefficient $\beta_3$ as a function of average effect size
$\langle \beta_3 \rangle$ and sample size of simulated
intermittent Gaussian distribution.

\end{document}